\documentclass
{JHEP3}

\usepackage{graphicx}
\usepackage{latexsym}
\usepackage{amssymb}
\usepackage{array}
\usepackage{slashed}

\def\0{\over } \def\1{\vec } \def\2{{\textstyle{1\over2}}} 
\def\3#1{{\bf{#1}}}
\def\4{{\textstyle{1\over4}}}
\def\5{\bar } 
\def\6{\partial }
\def\7#1{\slashed{#1}}
\def\8#1{{\textstyle{#1}}} 
\def\9{\underline }
\def\.{\cdot }
\def\^#1{\widehat{#1}}

\def\olrp{\hbox{\raisebox{2ex}{$\scriptstyle \leftrightarrow$}} \llap{$\partial$}}
\def\olp{\hbox{\raisebox{2ex}{$\scriptstyle \leftarrow$}} \llap{$\partial$}}

\def\({\left(} \def\){\right)} \def\<{\langle } \def\>{\rangle }
\def\[{\left[} \def\]{\right]}  
 
\def\pmbf#1{\setbox0=\hbox{${#1}$}
        \kern-.025em\copy0\kern-\wd0
        \kern.05em\copy0\kern-\wd0
        \kern-.025em\raise.0433em\box0 }

\def\be{\begin{equation}}
\def\ee{\end{equation}}

\newcommand{\bel}[1]{\begin{equation}\label{#1}}
\def\bea{\begin{eqnarray}}
\newcommand{\beal}[1]{\begin{eqnarray}\label{#1}}
\def\eea{\end{eqnarray}}
\def\nn{\nonumber\\ }

\preprint{
YITP-SB-07-20}

\title{Perturbative
Quantum Corrections to the\\ Supersymmetric CP$^1$ Kink with Twisted Mass}

\dedicated{Dedicated to the memory of Wolfgang Kummer } 

\author{Christoph Mayrhofer and Anton Rebhan\\
Institut f\"ur Theoretische Physik, Technische Universit\"at Wien,\\
  Wiedner Hauptstr. 8--10, A-1040 Vienna, Austria}

\author{Peter van Nieuwenhuizen and Robert Wimmer\\
C.N.Yang Institute for Theoretical Physics,
  SUNY at Stony Brook,\\ Stony Brook, NY 11794-3840, USA } 

\abstract{We present an explicit calculation of 
the one-loop quantum corrections to the mass
and the two central charges of the kink solution
of an $\mathcal N=(2,2)$ supersymmetric 
CP$^1$ model with
twisted mass, using 
supersymmetry preserving dimensional regularization 
adapted to solitons. We find that the quantum corrections of the mass
and one of the central
charges are nontrivial (but saturate the BPS bound), while the other
central charge receives no corrections. The nontrivial central charge
correction corresponds to a quantum anomaly, which in our scheme appears
as parity violation in the regulating extra dimension, and its magnitude
is in agreement with exact results obtained by Dorey on the basis
of a massive analog of mirror symmetry from a dual U(1) gauge theory,
confirming also the recent work by Shifman, Vainshtein, and Zwicky.}


\begin{document}

\section{Introduction}

Nonperturbative results on
supersymmetric (susy) theories
such as the famous Seiberg-Witten solution of $\mathcal N=2$ super-Yang-Mills
theories in 4 dimensions 
\cite{Seiberg:1994rs,Seiberg:1994aj,Alvarez-Gaume:1997mv}
rely crucially on the presence of solitons 
which saturate the Bogomolnyi bound at both the
classical level \cite{Bogomolny:1976de}
and at the quantum level \cite{Witten:1978mh}, namely when
equality of mass and central charge gives rise to multiplet shortening.
Results which are remarkably similar to
those of Seiberg and Witten have been obtained by Dorey \cite{Dorey:1998yh}
for a two-dimensional $\mathcal N=(2,2)$ U(1) gauge 
theory with $N$ chiral multiplets of equal charge and
twisted mass terms \cite{Hanany:1997vm}. At large gauge coupling
$e$ this theory
is in a Higgs phase whose low-energy limit is described by
a classically massive $\mathcal N=(2,2)$ CP$^{N-1}$ model
with BPS-saturated dyons which can carry both topological and
Noether charges. The dual (mirror) theory, where the
(dimensionful) gauge coupling $e$ is smaller than all other
mass scales, can be solved exactly and because the BPS spectrum
is independent of $e$ this
yields all-order results for the spectrum of the
CP$^{N-1}$ model as a function of the twisted masses.
Moreover these results turned out to be described by the same
elliptic curve that appeared in the Seiberg-Witten solution
of $\mathcal N=2$ gauge theories in four dimensions.

Recently, there has been renewed interest in this model since it
arises also as the effective field theory of so-called confined 
nonabelian monopoles, which reside within nonabelian flux tubes
(vortices) of $\mathcal N=2$ gauge theories with gauge
group SU($N$)$\times$U(1) and $N$ flavors 
\cite{Hanany:2003hp,Auzzi:2003fs,Auzzi:2003em,Shifman:2003uh,Shifman:2004dr,Hanany:2004ea,Gorsky:2004ad}. 
This connection in fact explained the observation of Ref.~\cite{Dorey:1998yh}
of a striking parallel between four-dimensional $\mathcal N=2$
super-Yang-Mills theory and the two-dimensional 
$\mathcal N=(2,2)$ CP$^{N-1}$ model, because the
four-dimensional Fayet-Iliopoulos parameter does not
enter the formulae for the spectrum of the BPS sector so that they
cover both the Higgs and the Coulomb branches.
The theories giving rise to confined monopoles in the Higgs phase
have an analytically accessible quasiclassical regime which
corresponds to twisted masses that are much larger than the scale of
the asymptotically free CP$^{N-1}$ model. There the coupling constant
of this effective theory is small and permits perturbative calculations.

A perturbative calculation of the quantum mass of the kink solution of the
$\mathcal N=(2,2)$ CP$^1$ model with twisted mass and a comparison
with the exact results obtained from the dual theory has been
made already in the original paper by Dorey \cite{Dorey:1998yh},
however without attempting accuracy beyond the logarithmic term
that shows up at one-loop order. As has been pointed out recently
by Shifman, Vainshtein and Zwicky \cite{Shifman:2006bs}, the
finite contribution that remains after absorbing the logarithmic
term into the renormalized coupling is associated with an anomalous
contribution to the central charge analogous to the one found some time ago
in ordinary susy kinks \cite{Shifman:1998zy,Goldhaber:2001rp,Rebhan:2002yw} and
which was subsequently located also in $\mathcal N=2$ super-Yang-Mills
theories both in its Coulomb phase \cite{Rebhan:2004vn} and its
Higgs phase \cite{Shifman:2004dr}.

In the present paper we complete the analysis begun by 
Dorey \cite{Dorey:1998yh}, namely a direct calculation of the quantum
mass of the CP$^1$ kink with twisted mass and also of the
central charges.
Such a calculation
involves the fluctuations of fermionic and bosonic fields in the
background of the kink which
despite isospectrality do not cancel due to a nonvanishing difference of
the spectral densities. The resulting expression is in fact
ultraviolet divergent and already in the minimally susy kink
model presents a number of intricacies and pitfalls. 
For example, a sharp energy cutoff
regularization incorrectly produces a null result for the finite terms of
the one-loop contribution to the mass \cite{Kaul:1983yt,Imbimbo:1984nq}
(and would do so also in the case of the susy CP$^1$ kink).
The inconsistency of this method and its
result with known results from the (nonsupersymmetric)
sine-Gordon model was pointed out in Ref.~\cite{Rebhan:1997iv}, which in 1997
reopened the issue of how to calculate quantum corrections for
susy solitons.
However, the alternative calculation presented in Ref.~\cite{Rebhan:1997iv}
which used mode number regularization in finite volumes
was polluted by boundary energy that occurs with periodic
or antiperiodic boundary conditions. In Ref.~\cite{Nastase:1998sy} this
issue was resolved (by use of topological boundary conditions) which
showed that the net quantum correction to the mass of a minimally susy kink is
negative. Since there appeared to be no quantum correction to the
central charge \cite{Imbimbo:1984nq}, this presented a problem with the
BPS bound, which
the authors of Ref.~\cite{Nastase:1998sy} conjectured to be the result
of a quantum anomaly. The latter was finally located by Shifman,
Vainshtein, and Voloshin \cite{Shifman:1998zy} as an anomalous
additive contribution to the central charge operator which
restores BPS saturation (which did not seem to be required
by standard multiplet shortening arguments \cite{Witten:1978mh}, but
could eventually be explained through the possibility of single-state
supermultiplets \cite{Losev:2000mm,Goldhaber:2000ab}). These
anomalous contributions to the central charge were confirmed in later
works, e.g.\ Ref.~\cite{Goldhaber:2001rp}, although by using dimensional
regularization methods Ref. \cite{Graham:1998qq} seemed to obtain 
the required finite
corrections to both mass and central charge without the need of an
anomalous contribution.

In Ref.~\cite{Rebhan:2002uk,Rebhan:2002yw,Rebhan:2004vn}, 
three of us performed one-loop calculations
using a variant of dimensional regularization in
the presence of solitons which embeds the solitons
in higher dimensions, from where susy-preserving 
dimensional reduction is possible.
This reproduces the correct results for
the quantum mass while indeed giving null results for
the original central charge operator. However, anomalous contributions
arise from nonvanishing bulk contributions to the momentum
density in the extra dimension which break reflection
invariance in the extra dimension, related to the fact
that fermionic zero modes turn into chiral domain wall fermions.
(Some additional issues arise for susy vortices in 2+1 dimensions
and the $\mathcal N=4$ monopole in 3+1 dimensions, see
Refs.~\cite{Rebhan:2003bu,Rebhan:2005yi,Rebhan:2006fg}.)
In the present paper we apply our scheme to the susy CP$^1$ model
with twisted mass term.

In superspace, the massless $\mathcal N=1$ 
CP$^1$ model in 4 dimensions or the $\mathcal N=(2,2)$
model in 2 dimensions can be written as
\be
\mathcal L=\int d^4\theta\,K(\Phi,\bar\Phi),\quad
K=r\ln(1+\bar\Phi\Phi)
\ee
with $\Phi$ a conventional chiral superfield, $\bar D_\alpha \Phi=0$.

In components, this reads, using the conventions of \cite{WB:SS},
\be\label{L4d}
\mathcal L=-{r\0\rho^2}
\biggl\{
\6_m\bar\phi \6^m \phi+i\bar\psi_{\dot\alpha}\bar\sigma^{m\dot\alpha\alpha}
(\6_m-{2\0\rho}\bar\phi_\alpha (\6_m\phi))\psi
+{1\02\rho^2}\psi\psi\bar\psi\bar\psi
\biggr\},\quad\rho\equiv 1+\phi^\dagger\phi,
\ee
where $m=0,\ldots 3$, and two of the $\6_m$ put to zero in
the dimensional reduction to 2 dimensions.
In 2 dimensions, the gauge coupling $g$ defined by $r={2\0g^2}$ 
is dimensionless
and its beta function is negative, so that the model is asymptotically free.
Correspondingly, at the quantum level this theory has a mass gap determined by
the renormalization group invariant scale $\Lambda$.

A classically massive version of the model in dimensions lower
then 4 which preserves the entire supersymmetry
can be obtained by introducing a background
gauge field with nonvanishing value in the components corresponding
to the dimensions eliminated in the reduction process,
\be\label{Vhat}
\6_m\to \6_m+i\hat V_m,\quad \hat V_m \6^m \Phi\equiv 0.
\ee
The mass terms provided by $\hat V_m=const.\not=0$ have been termed
twisted \cite{Hanany:1997vm}, 
because a gauge field strength superfield $\Sigma$ in two dimensions
is a twisted chiral superfield \cite{Gates:1984nk}, 
satisfying $\bar D_R\Sigma=D_L\Sigma=0$
instead of the conventional chiral constraint.

Dimensional reduction from 4 to 2 dimensions thus gives the
possibility for introducing two mass parameters, which can be
combined into one complex mass parameter $\tilde m=|m|e^{i\beta}$.
The phase $\beta$ corresponds to possible rotations in the
two dimensions used for the dimensional reduction, and it turns
out that because of the anomalous nature of the corresponding
U(1)$_A$ transformation its effect can be absorbed into a
$\theta$ term that can be added to the 2-dimensional Lagrangian.

The introduction of a mass term has the effect of providing the
(nonnegative) potential
term
\be\label{VCP1}
V={r\0\rho^2}|m|^2\phi^\dagger\phi={r|m|^2\phi^\dagger\phi\0(1+\phi^\dagger\phi)^2}
\ee
with zeros at $\phi=0$ and $\phi=\infty$, which correspond to the
north and south pole of the Riemann sphere, or CP$^1$,
obtained by compactifying
the complex plane parametrized by $\phi$.
The CP$^1$ kink is
the static field configuration which asymptotes to these two
different minima for left and right infinity.
We shall study its one-loop quantum corrections in the
perturbative regime provided by $m\gg\Lambda$, whereby the
coupling $g$ remains small for all energies.

\section{The model in 3 dimensions}\label{sec:3d}

Dimensional reduction of the $\mathcal N=(1,1)$ model (\ref{L4d})
in 4 dimensions with the modification (\ref{Vhat})
leads to the $\mathcal N=(2,2)$ sigma model with twisted mass term and the
CP$^1$ kink solution in 2 dimensions, but
in the following we shall reduce only from 4 to 3 dimensions,
keeping the extra dimension for the purpose of susy preserving
dimensional regularization by dimensional reduction. The dimension
needed to generate the twisted mass term as a vev of a (background) 
gauge field
component is thus compactified to vanishing size, but the other
extra dimension is kept. The
CP$^1$ kink of the 1+1-dimensional model
becomes a CP$^1$ domain wall (a line) in 2+1 dimensions.  

The action of the 2+1-dimensional model contains one complex scalar and
one complex 2-component spinor\footnote{Our conventions are 
$\{\gamma^\mu,
\gamma^\nu\}=2\eta^{\mu\nu}$ with $\eta^{\mu\nu}={\rm diag}(-1,+1,+1)$,
$\bar\psi=\psi^\dagger i\gamma^0$, thus $(\gamma^0)^2=-1$ and
$\gamma^{\mu\nu\rho}=-\epsilon^{\mu\nu\rho}$, $\gamma^{\mu\rho}=
-\epsilon^{\mu\rho\sigma}\gamma_\sigma$ with $\epsilon^{012}=+1$.
\label{footconv}} 
\bea\label{L3d}
\mathcal L&=&-{r\0\rho^2}\biggl[
\6_\mu \phi^\dagger \6^\mu\phi + m^2\phi^\dagger\phi + 
\bar\psi \gamma^\mu\6_\mu \psi + m \bar\psi\psi \left(1-{2\phi^\dagger\phi
\0\rho}\right)\nn
&&\qquad\quad -{2\0\rho}(\bar\psi\gamma^\mu\psi)(\phi^\dagger\6_\mu\phi)
-{1\0\rho^2}(\bar\psi\psi)(\bar\psi\psi) \biggr],
\qquad \mu=0,1,2,\quad\rho\equiv 1+\phi^\dagger\phi,
\eea
where we have arranged for standard kinetic and mass terms by choosing
a slightly unconventional ordering of Pauli matrices for
$\bar\sigma^M=(-\mathbf 1,-\sigma^1,-\sigma^3,-\sigma^2)$
in (\ref{L4d})
together with $\gamma^0=-i\sigma^2$. This fixes our conventions for
the $\gamma$ matrices in (\ref{L3d}) as
\be\label{3dgammas}
\gamma^0=-i\sigma_2, \quad \gamma^1=-\sigma_3, \quad \gamma^2=\sigma_1,
\ee
in agreement with the conventions used in our previous papers on susy kinks and their
embedding in 2+1 dimensions 
\cite{Rebhan:2002uk,Rebhan:2002yw,Goldhaber:2004kn} except for
the overall sign of $\gamma^1$. The direction of $x^2\equiv y$ 
will be our regulator dimension, and the two-dimensional
kink to be introduced shortly will depend only on $x^1\equiv x$. 
The reason for using $\sigma_3$ in $\gamma^1$ rather than $\sigma_1$ is that
this simplifies the fermionic fluctuation equations in the kink
background (see below). Note that in our conventions
the spinor components $\psi=\left( \psi^+ \atop \psi^-
\right)$ correspond to positive and negative
two-dimensional chirality with respect to the
regulating dimension $x^2$ (moving ``up'' and ``down'' the domain wall); 
the more conventional
left and right moving components of the final two-dimensional theory
are related to the former by $\psi^R=(\psi^++\psi^-)/\sqrt2$ and
$\psi^L=(\psi^+-\psi^-)/\sqrt2$.

The Lagrangian density (\ref{L3d})
is hermitian up to the antihermitian surface term \break
$\6_\mu \left({r\0\rho^2} \bar\psi \gamma^\mu \psi\right)$.
One can write this model in a $\psi$-$\bar\psi$ symmetric
way, or with the derivatives acting on $\bar\psi$ instead of $\psi$,
the only modifications being then, respectively,
\be\label{L3d2}
-{r\0\rho^2}\biggl[
\ldots +\2 \left(\bar\psi \gamma^\mu\olrp_\mu \psi\right) \ldots 
-{1\0\rho}(\bar\psi\gamma^\mu\psi)(\phi^\dagger\olrp_\mu\phi)
\ldots \biggr]
\ee
and
\be\label{L3d3}
-{r\0\rho^2}\biggl[
\ldots - \left(\bar\psi \gamma^\mu\olp_\mu \psi\right) \ldots 
+{2\0\rho}(\bar\psi\gamma^\mu\psi)(\phi^\dagger\olp_\mu\phi)
\ldots \biggr],
\ee
where it is understood that derivatives never act outside parentheses.

These actions are invariant under the following 
$\mathcal N=(2,2)$ rigid susy transformations
with two complex parameters $\epsilon^+$, $\epsilon^-$ with 
$\epsilon=\left(\epsilon^+ \atop \epsilon^-\right)$,
\bea\label{susytransf}
&&\delta\phi=\bar\epsilon\psi,\qquad \delta\phi^\dagger=\bar\psi\epsilon,\nn
&&\delta\psi=\gamma^\mu\6_\mu\phi\epsilon-m\phi\epsilon
+{2\phi^\dagger\0\rho}(\bar\epsilon\psi)\psi,\nn
&&\delta\bar\psi=-\bar\epsilon\gamma^\mu\6_\mu\phi^\dagger
-\bar\epsilon\phi^\dagger m+{2\phi\0\rho}(\bar\psi\epsilon)\bar\psi.
\eea

\section{The susy algebra}

The susy algebra on $\phi,\phi^\dagger,\psi$ has the following form
\def\ei{\epsilon_1} \def\ez{\epsilon_2}
\def\dei{\delta(\epsilon_1)} \def\dez{\delta(\epsilon_2)}
\def\eib{\bar\epsilon_1} \def\ezb{\bar\epsilon_2}
\def\deib{\delta(\bar\epsilon_1)} \def\dezb{\delta(\bar\epsilon_2)}
\bea\label{susyalgtransf}
&& [\deib,\dezb]=[\dei,\dez]=0,\nn
&& [\dei,\dezb] \left(\phi \atop \phi^\dagger \right)=
(\bar\epsilon_2\gamma^\mu\epsilon_1)\6_\mu \left(\phi \atop \phi^\dagger \right) 
\mp m(\bar\epsilon_2\epsilon_1) \left(\phi \atop \phi^\dagger \right)\nn
&&[\dei,\dezb] \psi = (\bar\epsilon_2\gamma^\mu\epsilon_1)\6_\mu \psi
- m(\bar\epsilon_2\epsilon_1) \psi
+\2 (\bar\epsilon_2\epsilon_1)F-\2(\bar\epsilon_2\gamma^\mu\epsilon_1)
\gamma_\mu F,
\eea
where $F$ is the complete field equation\footnote{%
Note that as in any nonlinear theory, the fermionic terms in the action
do not vanish on-shell; rather on-shell a term $(\bar\psi\psi)(\bar\psi\psi)$
remains.}
for $\psi$,
\be\label{F}
F=\7\6\psi+m\psi\left(1-{2\phi^\dagger\phi\0\rho}\right)
-{2\0\rho}\gamma^\mu\psi(\phi^\dagger\6_\mu\phi)
-{2\0\rho^2}(\bar\psi\psi)\psi.
\ee
(The susy commutator for $\bar\psi$ is easily derived by using
$\delta\bar\psi=\delta\psi^\dagger i\gamma^0$.)

The above algebra has the expected form of
\be
\{ Q,\bar Q \} = \gamma^\mu P_\mu + iZ
\ee
where $P_\mu$ is the antihermitian translation generated represented
by $\6_\mu$ in (\ref{susyalgtransf}) and 
$Z$ is the anti-hermitian central charge 
proportional to the unit matrix
which takes on the same value on both $\phi$ and $\psi$, because those
are in the same multiplet (and opposite value on the complex conjugate
multiplet with $\phi^\dagger$ and $\bar\psi$). 

The susy currents can be derived from the Noether method, by letting
the rigid $\epsilon$ become local. One finds
\be
j^\mu = {r\0\rho^2}\left[ \gamma^\rho (\6_\rho \phi^\dagger)
+ m \phi^\dagger \right]\gamma^\mu \psi,\quad
\bar j^\mu = {r\0\rho^2} \bar\psi \gamma^\mu
\left[ \gamma^\rho (\6_\rho \phi)- m \phi \right].
\ee
One may check that $\delta_{\bar\epsilon}\phi=[-i\bar\epsilon Q,\phi]$,
$\delta_{\bar\epsilon}\bar\psi=[-i\bar\epsilon Q,\bar\psi]$ and
$\delta_{\epsilon}\bar\psi=[-i\bar Q\epsilon,\bar\psi]$
with $Q=\int j^0 dx\,dy$
reproduce the transformation rules with canonical conjugate momenta
\be\label{conjmoms}
p(\phi)={r\0\rho^2}\dot\phi^\dagger+
{2r\0\rho^3}(\bar\psi\gamma^0\psi)\phi^\dagger,
\quad p(\phi^\dagger)={r\0\rho^2}\dot\phi,
\quad
p(\psi)={r\0\rho^2}\bar\psi\gamma^0
\ee
with $\{p(\psi)(t,\mathbf x),\psi(t,\mathbf y)\}=
-i\delta^2(\mathbf x-\mathbf y)$.
(No Dirac brackets are necessary if one uses (\ref{L3d}) and
replaces $\bar\psi$ by $p(\psi)$
as indicated, but note that (\ref{conjmoms}) implies that
$p(\phi^\dagger)$ is not equal to $(p(\phi))^\dagger$ if one uses
naive hermitian conjugation.)

\section{Classical CP$^1$ kink and domain line}

The classical kink (domain wall)
solution interpolating between the two minima 
$\phi=0$ and $\phi=\infty$ of the potential (\ref{VCP1}) for the bosonic fields
is most easily found
by completing squares in the bosonic part of the classical
Hamiltonian density. Assuming dependence of $\phi$ on
only the $x$ coordinate, we have
\be
\mathcal H={r\0\rho^2} (\6_x \phi^\dagger-m\phi^\dagger)
 (\6_x \phi-m\phi)+\6_x\left(-rm\0\rho\right).
\ee
So the classical kink solution and its mass are
\be\label{phiK}
\phi_K=e^{m(x-x_0)+i\alpha},\qquad M_{cl}=rm.
\ee
There are two real moduli, $x_0$ and $\alpha$, and correspondingly
two real (one complex) zero modes, see (\ref{g0x}).

The classical kink solution preserves one half of susy: from
(\ref{susytransf}) with $\delta\psi=0$  and 
$\gamma^1=\left( {-1 \atop 0} {0 \atop 1} \right)$
we see that the remaining susy is given by $\epsilon=\left(0
\atop \epsilon^-\right)$. 
The broken susy with $\epsilon=\left(\epsilon^+ \atop 0\right)$
produces the fermionic zero mode 
\be
\psi \sim \phi_K \left(\epsilon^+ \atop 0 \right).
\ee

Since the generators of the preserved susy are $\bar Q\epsilon=
-i(Q^+)^\dagger\epsilon^-$ and
$\bar\epsilon Q=i(\epsilon^-)^\dagger Q^+$, we see that
$Q^+$ and $(Q^+)^\dagger$ preserve the solitonic ground
state $|sol\>$. BPS saturation at the quantum level thus requires
\be\label{BPSsat}
\< sol | \{ Q^+,(Q^+)^\dagger \} | sol \> = 0.
\ee
This implies that $\int(T^0{}_0+T^0{}_2)dx\,dy$ should vanish.
In the classical 2-dimensional model, $T^0{}_2$ is a regularized central
charge density, and $\zeta^0$ a second one.
To evaluate them at the quantum level, we need to obtain the
currents $T^\mu{}_\nu$ and $\zeta^\mu$.

\section{Energy momentum tensor and central charge currents}

The variation $\delta(\bar\epsilon)j^\mu$ vanishes, as one easily
checks, but for $\delta(\epsilon)j^\mu$ we find, after tedious but
straightforward algebra, using Fierz rearrangements but never
discarding terms that are total derivatives, the following results
\be
\delta(\epsilon)j^\mu=T^\mu{}_\nu \gamma^\nu\epsilon+\zeta^\mu\epsilon
\qquad (\mu,\nu=0,1,2)
\ee
where
\bea
T^\mu{}_\nu &=&
{r\0\rho^2} \biggl[ \6^\mu \phi^\dagger \6_\nu\phi + \6_\nu \phi^\dagger
\6^\mu \phi^\dagger - \delta^\mu_\nu (\6^\lambda\phi^\dagger\6_\lambda\phi
+m^2\phi^\dagger\phi) -\2 (\6^\mu \bar\psi)\gamma_\nu \psi
-\2 (\6_\nu \bar\psi)\gamma^\mu \psi \nn
&&\quad+{1\0\rho}(\6^\mu\phi^\dagger)\phi\bar\psi\gamma_\nu\psi
+{1\0\rho}(\6_\nu\phi^\dagger)\phi\bar\psi\gamma^\mu\psi
-\delta^\mu_\nu {1\0\rho^2}(\bar\psi\psi)(\bar\psi\psi)
-\2\delta^\mu_\nu \bar F\psi
\nn
&&\quad +\epsilon^\mu{}_\nu{}^\lambda\left\{
m\6_\lambda(\phi^\dagger\phi)-{m\02}\bar\psi\gamma_\lambda\psi
\left(1-{2\phi^\dagger\phi\0\rho}\right)
+\2(\6_\lambda\bar\psi)\psi-{(\6_\lambda\phi^\dagger)\phi\0\rho}\bar\psi\psi
\right\}\biggr]
\eea
Here $\bar F$ is the complete field equation of $\bar\psi$,
\be
\bar F=-\6_\mu\bar\psi \gamma^\mu
+m\bar\psi\left(1-{2\phi^\dagger\phi\0\rho}\right)
+{2\0\rho}\bar\psi(\7\6\phi^\dagger)\phi
-{2\0\rho^2}(\bar\psi\psi)\bar\psi.
\ee

On-shell $T^\mu{}_\nu$ is not symmetric, nor should it be symmetric,
for two reasons: it is not the gravitational stress tensor, and it
may contain total derivatives which are antisymmetric in $\mu,\nu$.
These total derivatives will contribute to the central charge.
In order to obtain a $T^\mu{}_\nu$ which is symmetric up to
total derivatives (and in which $\psi$ and $\bar\psi$ appear on
equal footing) one can proceed in two ways:
either one adds $\deib(\bar j^\mu\epsilon_2)$ to $\dez \bar\epsilon_1 j^\mu$
(which both come from $[\bar\epsilon_1 Q,\bar Q\epsilon_2]$)
and divides by 2,
or one partially integrates various terms in $T^\mu{}_\nu$, keeping
track of total derivatives.
The result is the same and reads
\bea
T^\mu{}_\nu &=&
{r\0\rho^2} \biggl[ \6^\mu \phi^\dagger \6_\nu\phi + \6_\nu \phi^\dagger
\6^\mu \phi^\dagger - \delta^\mu_\nu (\6^\lambda\phi^\dagger\6_\lambda\phi
+m^2\phi^\dagger\phi) +\4 (\bar\psi\gamma^\mu \olrp_\nu\psi )
+\4 (\bar\psi\gamma_\nu \olrp^\mu\psi ) \nn
&&\quad-{1\02\rho}(\phi^\dagger\olrp^\mu\phi)\bar\psi\gamma_\nu\psi
-{1\02\rho}(\phi^\dagger\olrp_\nu \phi)\bar\psi\gamma^\mu\psi
-\delta^\mu_\nu {1\0\rho^2}(\bar\psi\psi)(\bar\psi\psi)
-\4\delta^\mu_\nu (\bar F\psi+\bar\psi F)\biggr]
\nn
&&\quad +r\epsilon^\mu{}_\nu{}^\lambda \6_\lambda\left\{
-{m\0\rho}+{1\04\rho^2}\bar\psi\psi
\right\}
\eea
The first two lines now correspond to the gravitational stress tensor,
where all terms with $\delta^\mu_\nu$ can be written as $\delta^\mu_\nu
\mathcal L$ with $\mathcal L$ from (\ref{L3d2})
and the last term, which is a total derivative, is the
only one antisymmetric in $\mu,\nu$. Note that although the various
ways of writing the action, eqs. (\ref{L3d})-(\ref{L3d3}), differ by
total derivatives, there is no ambiguity in the total derivatives
in this $T^\mu{}_\nu$, because it is by definition due to the susy
variation of the susy current $j^\mu$, and the latter is
unambiguous.\footnote{We exclude topological terms in the susy current
because they would lead to modifications of the susy transformations
at the boundary.}

The central charge current $\zeta^\mu$ is found to be given by
\bea
\zeta^\mu &=& \epsilon^{\mu\nu\lambda} {\6_\nu\phi^\dagger \6_\lambda\phi\0
\rho^2}
+{m\0\rho^2}(\phi^\dagger\olrp^\mu\phi)-{m\02\rho^2}(\bar\psi\gamma^\mu
\psi)\left(1-{2\phi^\dagger\phi\0\rho}\right)\nn
&&+{1\0\rho^3}(\6_\lambda\phi^\dagger)\phi \bar\psi \gamma^\lambda
\gamma^\mu\psi - {1\02\rho^2}(\6_\lambda\bar\psi)\gamma^\lambda
\gamma^\mu\psi.
\eea
Again we can either partially integrate half of the last term,
or subtract $\tilde\zeta^\mu$ (and divide by 2), where
$\delta(\bar\epsilon)\bar j^\mu=\tilde T^\mu{}_\nu(-\bar\epsilon\gamma^\nu)
+\tilde\zeta^\mu\bar\epsilon$. The result is the same on-shell and reads
\be\label{zetamu}
\zeta^\mu=\epsilon^{\mu\nu\lambda} {\6_\nu\phi^\dagger \6_\lambda\phi
\0\rho^2} +{m\0\rho^2}
\left[ (\phi^\dagger\olrp^\mu\phi-\bar\psi\gamma^\mu\psi
\left(1-{2\phi^\dagger\phi\0\rho}\right)\right]+
{1\02\rho^2}\bar F\gamma^\mu\psi ,
\ee
where we used that $(\bar\psi\psi)(\bar\psi\gamma^\mu\psi)=0$.

\section{Quantization}

For the evaluation of one-loop quantum corrections we need to obtain
the fluctuation equations in the CP$^1$ kink background $\phi_K$.

The fermionic fluctuations satisfy the field equation
(\ref{F}), and to linear order in $\psi$
with $\phi=\phi_K$ one has
\be
\7\6\,\psi+m\psi\left(1-{2\phi^\dagger\phi\0\rho}\right)
-{2\0\rho}\gamma^\mu\psi(\phi_K^\dagger\6_\mu\phi_K)=0.
\ee
Using the explicit form of the kink solution (\ref{phiK}), 
with $x_0=0$ and $\alpha=0$ for simplicity, 
and our representation of the $\gamma$ matrices as given
in (\ref{3dgammas}) we obtain
\be\label{fermfluc}
\left(\begin{array}{cc}
\tilde L & -\6_0+\6_y \\
\6_0+\6_y & L
\end{array}\right)
\left(\begin{array}{c}
\psi^+ \\ \psi^-\end{array}\right)
=0,\qquad
\begin{array}{l}
\tilde L = -\6_x+m, \\
L = \6_x+m-4me^{2mx}/(1+e^{2mx}).
\end{array}
\ee
With respect to an inner product defined by $(\lambda,\chi)=\int
{1\0\rho^2}\lambda^*\chi\, dx$, the operator $\tilde L$ is the adjoint of
$L$, $(\lambda,L\chi)=(\tilde L\lambda,\chi)$ up to surface terms.
Iterating (\ref{fermfluc}) yields
\bea
&&(L\tilde L-\6_y^2+\6_0^2)\psi^+=0,\\
&&(\tilde L L-\6_y^2+\6_0^2)\psi^-=0.
\eea
The operators $L\tilde L$ and $\tilde L L$ are selfadjoint without
surface terms, so they yield a complete set of eigenfunctions.
Let $\varphi_k(x)$ be a solution of
\be
L\tilde L\varphi_k=\omega_k^2\varphi_k
\qquad \mbox{with $\omega_k^2=k^2+m^2$},
\ee
and let
\be\label{sk}
s_k={1\0\omega_k}\tilde L \varphi_k.
\ee
Then in second quantization\goodbreak
\bea\label{psimodeexp}
\left( \psi^+ \atop \psi^- \right)&=&{1\0\sqrt{r}}
\int{dk\,d^\epsilon\ell \0 (2\pi)^{(1+\epsilon)/2}}
{1\0\sqrt{2\omega}} \biggl[
\alpha_{k\ell} \left( \sqrt{\omega+\ell}\,\varphi_k(x) 
\atop \sqrt{\omega-\ell}\, i s_k(x)\right)
e^{i\ell y-i\omega t}\nn
&&\qquad\qquad\qquad\qquad\qquad
+\beta_{k\ell}^\dagger \left( \sqrt{\omega+\ell}\,\varphi_k^*(x)
\atop -\sqrt{\omega-\ell}\, i s_k^*(x)\right)
e^{-i\ell y+i\omega t}\,\biggr]\nn&& +
{1\0\sqrt{r}} \int {d^\epsilon\ell \0 (2\pi)^{\epsilon/2}}
\gamma_\ell \left( \varphi_0(x) \atop 0 \right) e^{i\ell(y-t)}\,,
\eea
where $\left( \psi^+ \atop \psi^- \right)$ satisfies (\ref{fermfluc}),
and $\omega^2\equiv k^2+\ell^2+m^2$. Here $\ell$ is the momentum
component along the domain wall, and we have already indicated that dimensional
regularization by dimensional reduction will eventually be performed
by sending $\epsilon$ from 1 to 0.
The last term is due to the fermionic zero mode, which in dimensions
larger than 2 turns into a continuum of 
massless modes localized along the domain
line and with definite chirality with respect to the latter.
The correct normalization of this term can be obtained by
taking the formal limit $\omega_k\to0$ in the nonzero mode terms and
combining the terms with $\ell>0$ and $\ell<0$ into one term
with $-\infty<\ell<\infty$, setting $\{\gamma_\ell,\gamma_{\ell'}^\dagger\}
=\delta(\ell-\ell')$.
Note that $\gamma_\ell$ ($\gamma_\ell^\dagger$) have the meaning
of annihilation (creation) operators only for $\ell>0$ and that for $\ell<0$
this is to be reversed. As (\ref{psimodeexp}) shows, the positive frequency
modes have momentum in positive $y$-direction only, so that there
is a breaking of parity invariance with respect to the regulator
dimension. The opposite breaking would have taken place with
the choice $\gamma^2=-\sigma^1$, which gives a nonequivalent second
representation of the Clifford algebra in 3 dimensions. 

The bosonic fluctuations $\eta$ are obtained from $\phi=\phi_K+\eta$, and
after some work one finds for their linearized field equations the
same result as for $\psi^+$,
\be
 (L\tilde L-\6_y^2+\6_0^2)\eta=0.
\ee
To solve this equation we first look at its behaviour at large $|x|$,
where $L\tilde L\to -\6_x^2+4m\6_x-3m^2$ as $x\to+\infty$ and
$L\tilde L\to -\6_x^2+m^2$ as $x\to-\infty$. We set then
\be
\eta(x)=(1+e^{2mx})g(x)
\ee
and find for $g(x)$ the differential equation
\be
\left[-\6_x^2+m^2-{2m^2\0\cosh^2(mx)}\right]g=\omega_k^2\,g
\ee
This is the $l=1$ case of the sequence of operators
\be \mathcal O_l=A_l^\dagger A_l=
-\6_z^2+l^2-{l(l+1)\0\cosh^2 z}
\ee
with $A_l=\6_z+l\tanh z$ and $A^\dagger=-\6_z+l\tanh z$, where $z=mx$.
For $l=1$, this system, which also appears in the 2-dimensional
sine-Gordon model\footnote{The sine-Gordon model also appears in the dual
formulation of the CP$^1$ model 
\cite{Fateev:1979dc,Fendley:1992dm,Hori:2000kt}}, 
contains one zero mode, no bound state, and
a continuum of solutions, given respectively by
\bea\label{g0x}
g_0(x)&=& \sqrt{m\02}{1\0\cosh(mx)},\\
\label{gkx}
g_k(x)&=& {1\0\sqrt{2\pi}}{-ik+m\tanh(mx)\0\omega_k}e^{ikx}.
\eea
Note that $g_0$ corresponds to $\varphi_0(x)=
\rho_K(x)g_0(x)=\sqrt{2m}e^{mx}$ 
which is indeed proportional to the function arising from differentiating
$\phi_K$ in (\ref{phiK}) with respect to either of the
moduli $x_0$ or $\alpha$.

Then in second quantization
\bea
\eta(t,x,y)&=&{1\0\sqrt{r}}
\int{dk\,d^\epsilon\ell \0 (2\pi)^{(1+\epsilon)/2}}
{1\0\sqrt{2\omega}} \left[ a_{kl} \varphi_k(x) e^{i\ell y-i\omega t}
+b_{kl}^\dagger \varphi_k^*(x) e^{-i\ell y+i\omega t}\right]\nn
&&+{1\0\sqrt{r}}\int {d^\epsilon\ell \0 (2\pi)^{\epsilon/2}}
{1\0\sqrt{2|\ell|}}\left[c_\ell \varphi_0(x)e^{i\ell y-i|\ell|t}
+d_\ell^\dagger \varphi_0(x)e^{-i\ell y+i|\ell|t} \right],
\eea
with $\omega^2=\omega_k^2+\ell^2=
k^2+\ell^2+m^2$.
Given the normalization of $g_k(x)$ to plane waves at infinity, we
have the following orthonormality relations 
\be
\int {dx\0\rho^2(x)}\varphi_0^2(x)=1,\quad
\int {dx\0\rho^2(x)}\varphi_k^*(x)\varphi_{k'}(x)=\delta(k-k'),\quad
\int {dx\0\rho^2(x)}\varphi_0(x)\varphi_k(x)=0.
\ee

We shall also need the difference of
the spectral densities associated with the continuum solutions
$\varphi_k$ and $s_k$,
which is defined by
\be
\Delta\sigma(k)=\int {dx\0\rho^2(x)} \left(
|\varphi_k(x)|^2-|s_k(x)|^2 \right).
\ee
Using $s_k={1\0\omega_k}\tilde L \varphi_k$ and partially integrating,
only a surface term is left, and we find
\be\label{Deltarho}
\Delta\sigma(k)={\varphi_k^* \tilde L \varphi\0\omega_k^2 \rho^2(x)}
\Big|_{x=-\infty}^{x=\infty} = {-2m\0\omega_k^2} = {-2m\0k^2+m^2}.
\ee
This result agrees with the analysis of Ref.~\cite{Dorey:1998yh}, where
a nonlinear transformation of the fluctuating fields was employed
that simplified the fluctuation equations, but which corresponds
to a reparametrization of the fields that cannot be used in
perturbation theory about the topologically trivial vacuum, where
the renormalization of the model is to be fixed (one of the real
fields has no kinetic term in the vacuum). 
Our approach thus has the advantage of not having to combine results
from calculations using different parametrizations of the target space,
but a posteriori we find that no mistake would have been made
by doing so.

\section{The mass of the CP$^1$  kink}

The classical kink mass $M_{cl}=rm$ gets quantum corrections from
the zero point energies of the fluctuating fields and from renormalization,
\be
M^{(1)}=\int dx\, \<T_{00}^{(1)}\> + {\Delta r\0r} M_{cl}
\ee
where the subscript (1) refers to one-loop order contributions
and where we have anticipated that only $r$ and not $m$ gets renormalized
in our model, which is in fact true to all orders in perturbation theory
\cite{Alvarez-Gaume:1980dk}.

The one-loop renormalization $r_0=r+\Delta r$ of the coupling constant
$r\equiv 2/g^2$ can be obtained from the scalar self energy corrections
(or equivalently from the fermionic ones) in the trivial vacuum. 
Imposing the renormalization
condition that they vanish fixes $\Delta r$,
\be
\includegraphics[bb=150 725 440 775,clip,scale=0.9]{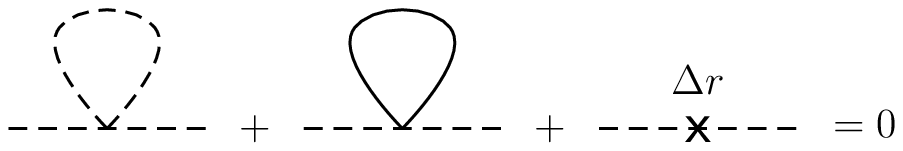}
\ee
By straightforward calculation we find
\be
\includegraphics[bb=160 725 325 775,clip,scale=0.7]{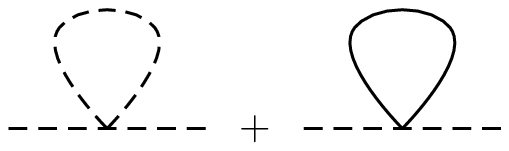}
\raisebox{5mm}{$\displaystyle =2\int
{d^{2+\epsilon}k\0(2\pi)^{2+\epsilon}}
{p^2+m^2-(k^2+m^2)\0k^2+m^2-i\epsilon}
$}
\ee
The integral with $-(k^2+m^2)$ in the numerator vanishes in dimensional
regularization, whereas the terms with $p^2+m^2$ can be canceled
by a counterterm $\Delta r$, leaving $m$ unrenormalized. This leads to
\be\label{Deltar}
\Delta r=\int {dk\,d^\epsilon\ell\0(2\pi)^{1+\epsilon}}{1\0\omega},
\quad \omega=\sqrt{k^2+\ell^2+m^2},
\ee
where the sign of this result corresponds to the well-known asymptotic
freedom of this model.

The bulk contributions to the mass are given by
\be
\<T_{00}^{(1)}\>=
\left\< {r\0\rho^2} \left( \6_0\phi^\dagger\6_0\phi+
\6_k\phi^\dagger\6_k\phi+m^2 \phi^\dagger\phi-\2\bar\psi\gamma^0\olrp_0\psi
\right)\right\>
\ee
where we dropped the terms with the fermionic field equations.
Rewriting the bosonic terms in this expression as 
$(2r\rho^{-2}\6_0\phi^\dagger\6_0\phi
-\mathcal L)$ and using that 
for any action
$\<\mathcal L_{ferm.}^{(2)}\>=0$ but
$\mathcal L_{bos}^{(2)}=0$ only up to boundary terms, we can
recast $\<T_{00}^{(1)}\>$ as follows
\be
\<T_{00}^{(1)}\>={r\0\rho^2}
\left\<2\6_0\eta^\dagger\6_0\eta-\bar\psi\gamma^0\6_0\psi\right\>
+\,\mbox{total derivatives}
\ee
The total derivatives are given by
\be
r\6_x\left[ {m\phi_K^2\0\rho_K^3}(\eta+\eta^\dagger)^2 \right]
-r \6_\mu \left[ {\eta\6^\mu\eta^\dagger\0\rho_K^2} \right],
\ee
but they do not contribute to the energy. 
(The propagator $\<\eta\eta^\dagger\>$ is proportional to $\rho_K^2$, and
the derivatives of $\rho_K$ in the second term cancel the first term.
One is left with a $\rho$-independent term with a derivative
on the distorted plane wave, and this term is the same at plus and minus
infinity.) 

Substituting the mode expansion of $\eta$ and $\psi$ yields
\bea\label{M1bulk}
M^{(1)}_{\rm bulk}&=&\int dx\, \<T_{00}^{(1)}\>
=\int {dx\0\rho^2} \!\int\! {dk\,d^\epsilon\ell\0(2\pi)^{1+\epsilon}2\omega}
\left[2\omega^2|\varphi_k|^2-\omega
\left\{(\omega+\ell)|\varphi_k|^2+(\omega-\ell)|s_k|^2\right\}
\right]\nn
&=&\int {dx\0\rho^2} \!\int\! {dk\,d^\epsilon\ell\0(2\pi)^{1+\epsilon}}
{\omega\02}\left( |\varphi_k(x)|^2 - |s_k(x)|^2 \right)
=-\int\! {dk\,d^\epsilon\ell\0(2\pi)^{1+\epsilon}}{m\omega\0\omega_k^2},
\eea
where we used the expression for the difference of
spectral densities obtained in eq.~(\ref{Deltarho}).
We see here clearly the sums over zero-point energies
($\sum\hbar\omega$ for complex scalars, $-\sum\hbar\omega$ for complex
fermions) and that despite of supersymmetry and isospectrality
there is a net contribution due to a difference of the
spectral density of the continuum modes. This contribution is
in fact ultraviolet divergent and becomes finite upon combining
it with the counterterm $\Delta r\, m$. Using the integral
representation of $\Delta r$ of eq.~(\ref{Deltar}) the total
mass correction is given by
\bea\label{M1}
M^{(1)}&=&m\int\! {dk\,d^\epsilon\ell\0(2\pi)^{1+\epsilon}}
\left({-m\omega\0\omega_k^2}+{m\0\omega}\right)
=-m\int\! {dk\,d^\epsilon\ell\0(2\pi)^{1+\epsilon}}
{\ell^2\0\omega\omega_k^2}\nn&&=
-{4\01+\epsilon}{\Gamma(1-\epsilon/2)\0(4\pi)^{1-\epsilon/2}}m^{1+\epsilon/2}
=-{m\0\pi}+O(\epsilon),
\eea
which is finite for all $\epsilon<2$. For $\epsilon=0$ one obtains
the nonvanishing correction $M^{(1)}=-m/\pi$ for the mass of the
susy CP$^1$ kink; for $\epsilon=1$ the result corresponds to
the mass per unit length of the domain line and then reads
$-m^2/(4\pi)$. Both results are precisely twice the universal\footnote{Because
of supersymmetry the difference in the spectral densities which
is responsible for the nonzero result is determined
by the asymptotic values of the fermion mass and does not depend 
on other details of the potential \cite{Rebhan:1997iv,Nastase:1998sy}.}
amount
one finds for minimally supersymmetric 1+1-dimensional
kinks and 2+1-dimensional
domain lines, respectively, provided the latter are renormalized
in a minimal scheme \cite{Rebhan:2002uk}.
By contrast, ordinary $\mathcal N=2$ susy kinks
in Landau-Ginzburg type models lead to complete
cancellations of the quantum corrections  \cite{Nastase:1998sy} instead
of the doubling we found here for the $\mathcal N=2$ nonlinear
sigma model with twisted mass term.

Next we shall consider the quantum corrections to the central charges,
which have to involve the same finite correction in order that
BPS saturation holds. This will moreover show that 
these finite corrections are associated with an anomaly.

\section{The central charges}

The central charge responsible for the saturation of the BPS bound
is associated with $T^0{}_2$ of the 3-dimensional model, as follows from
(\ref{BPSsat}). Its evaluation now involves bulk contributions,
boundary terms, and a renormalization term,
\bea
T^0{}_2&=&{r\0\rho^2}\left[
-\6_0\phi^\dagger\6_2\phi-\6_2\phi^\dagger\6_0\phi
+\4\bar\psi\gamma^0\olrp_2 \psi-\4\bar\psi\gamma^2\olrp_0 \psi
\right]\nn
&&+r\6_x\left({m\0\rho}-{\bar\psi\psi\04\rho^2}\right)+
\Delta r\, \6_x{m\0\rho}.
\eea
As is usual for central charge corrections in susy models 
\cite{Imbimbo:1984nq}, loop corrections from the bosonic surface terms
cancel the renormalization term exactly,
\be
r\left\< {m\0\rho} \right\>\Big|_{-\infty}^\infty
=r{m\0\rho^3}2\phi^\dagger\<\eta\eta^\dagger\>\phi\Big|_{-\infty}^\infty
=\int\! {dk\,d^\epsilon\ell\0(2\pi)^{1+\epsilon}}{m\0\omega}
=m\Delta r = -\Delta r{m\0\rho}\Big|_{-\infty}^\infty.
\ee

Quite unusually, the fermionic surface term does contribute and
is even divergent,
\bea\label{fermsurfterm}
-{r\04\rho^2}\<\bar\psi\psi\>\Big|_{-\infty}^\infty &=&
{1\0\rho^2}\int\! {dk\,d^\epsilon\ell\0(2\pi)^{1+\epsilon}}{\omega_k\08\omega}
(\varphi_k s_k^*+s_k\varphi_k^*)\Big|_{-\infty}^\infty \nn
&=&
{1\0\rho^2}\int\! {dk\,d^\epsilon\ell\0(2\pi)^{1+\epsilon}}{1\08\omega}
(-2\rho \6_x \rho+2m\rho^2)\Big|_{-\infty}^\infty
=-{m\02}\int\! {dk\,d^\epsilon\ell\0(2\pi)^{1+\epsilon}}{1\0\omega}.
\eea

The bosonic bulk terms vanish since they are odd in $\ell$, but
the fermionic bulk terms do contribute a nonvanishing momentum
density along the domain line as follows,
\bea
&&{-i\02}r\int{dx\0\rho^2} \!\int\! {dk\,d^\epsilon\ell\0(2\pi)^{1+\epsilon}}
\left\<
(\psi^+)^\dagger (\6_2-\6_0)\psi^++(\psi^-)^\dagger
(\6_2+\6_0)\psi^-\right\> \nn
&&=-\2\int{dx\0\rho^2} 
\!\int\! {dk\,d^\epsilon\ell\0(2\pi)^{1+\epsilon}2\omega}
(\omega^2+\ell^2)(|\varphi_k|^2-|s_k|^2)
=\int\! {dk\,d^\epsilon\ell\0(2\pi)^{1+\epsilon}}{(\omega^2+\ell^2)m\0
2\omega\omega_k^2},
\eea
where once again (\ref{Deltarho}) has been used.
The total central charge $Z_1$ is finite and given by
\be\label{Z1}
Z_1^{(1)}=m\int\! {dk\,d^\epsilon\ell\0(2\pi)^{1+\epsilon}}
{\omega^2+\ell^2-\omega_k^2\02\omega\omega_k^2}=
m\int\! {dk\,d^\epsilon\ell\0(2\pi)^{1+\epsilon}}
{\ell^2\0\omega\omega_k^2}.
\ee
Comparing with (\ref{M1}), we see that BPS saturation holds,
$M^{(1)}+Z_1^{(1)}=0$.

The other central charge is $Z_2=\int \zeta^0 dx$, where
according to (\ref{zetamu})
\be\label{zeta0}
\zeta^0=\epsilon^{0\nu\lambda} {\6_\nu\phi^\dagger \6_\lambda\phi
\0\rho^2} +{m\0\rho^2}
\left[ (\phi^\dagger\olrp^0\phi-\bar\psi\gamma^0\psi
\left(1-{2\phi^\dagger\phi\0\rho}\right)\right].
\ee
It generates the $m$-dependent terms in (\ref{susyalgtransf}).
Considering one-loop corrections, one finds that
in momentum space the first term gives rise to an expression 
which is odd in $\ell$ and thus gives
no contribution. The second term gives rise to
\be
{2m\0\rho^2}\<\eta^\dagger \6_0 \eta\>-{4m\0\rho^3}\phi^\dagger
\<\6_0\eta \eta^\dagger\>\phi
\ee
and these terms vanish because they are independent of the
extra momentum $\ell$, leading to a scaleless integral
which is zero in dimensional regularization.
The contribution from the
third term (\ref{zeta0}) is also $\ell$-independent, because
the $\ell$ in $(\omega+\ell)|\varphi_k|^2$ and $(\omega-\ell)|s_k|^2$
(produced by the mode expansion (\ref{psimodeexp}))
cancels by symmetric integration, after which the remaining $\omega$
cancels the energy denominator ${1\02\omega}$. Hence, the second
central charge does not receive any one-loop corrections.

\section{Discussion and conclusions}

As mentioned in the Introduction, an exact result for
the central charge of the quantum CP$^1$ kink in the
nonlinear sigma model with a twisted mass term has been obtained
by Dorey \cite{Dorey:1998yh} in a generalization of results
of Hanany and Hori \cite{Hanany:1997vm}, 
which for the kink configuration
reads
\be\label{ZD}
\<Z\>={1\0\pi} 
\sqrt{\tilde m^2+4\tilde\Lambda^2}+{\tilde m\02}\ln{\tilde m-\sqrt{\tilde m^2+4\tilde\Lambda^2}\0
\tilde m+\sqrt{\tilde m^2+4\tilde\Lambda^2}},
\ee
where $\tilde m=m e^{i\beta}$ is the complex twisted mass parameter
mentioned in the Introduction,
and $\tilde\Lambda$ is the renormalization-group
invariant scale of the model, which is real in the absence of
a theta term. 
With the
identification $r=2g^{-2}={1\02\pi}\ln(m^2/\tilde\Lambda^2)$, the
weak-coupling limit of (\ref{ZD}) corresponds to $m\gg \tilde\Lambda$,
and expanding (\ref{ZD}) in this limit yields
\be\label{Z1D}
|\<Z\>|=\left|\tilde m{1\02\pi}\ln\left(-{\tilde m^2\0\tilde\Lambda^2}\right)
-{\tilde m\0\pi}\right|.
\ee
Identifying our (real) mass parameter $m$ with $|\tilde m|$ and
choosing $|\beta|=\pi/2$ such that the logarithm is real,
(\ref{Z1D}) reduces to $|\<Z\>|=rm-m/\pi$,
in agreement with our real results for the one-loop correction
of mass and central charge, (\ref{M1}) and (\ref{Z1}).\footnote{A
possible theta angle appears in the exact
result (\ref{ZD}) of Ref.~\cite{Dorey:1998yh}
as a phase of $\tilde\Lambda$ in such a way
that the phase of $\tilde m$ can be absorbed by a change of $\theta$.
However, using our scheme of dimensional regularization by
embedding the kink in one higher dimension we have
to restrict ourselves to $\theta=0$.} 

The possible imaginary part in $\<Z\>$ has to be identified
with the second central charge, $Z_2=\int dx\,\zeta^0$, considered above,
which contains the Noether charge density for the global U(1) symmetry
$\psi\to e^{i\lambda}\psi$, $\phi\to e^{i\lambda}\phi$ of (\ref{L3d}).
Besides the ``purely magnetic'' kink (\ref{phiK}), this model
also contains dyons, which are given by replacing
the constant $\alpha$ by $\alpha(t)=\omega t$ in (\ref{phiK}),
where at the quantum level $\omega$ is quantized by a Bohr-Sommerfeld
condition. In the above, we have considered a purely magnetic kink,
but the exact result (\ref{ZD}) shows that for general $\beta$ (and
also for general $\theta$) one has dyonic states.
In our calculation we have not obtained a contribution to $Z_2$ so
that our result corresponds to a purely imaginary $\tilde m$ in (\ref{ZD}).
Such a null result for the U(1) charge of the solitonic ground
state does not contradict the fact that the latter should
be defined as carrying fractional fermion number \cite{Jackiw:1976fn}
because of the presence of
fermionic zero modes. Indeed, the U(1) charge associated with
the fermionic zero mode vanishes:
\be
r\int {dx\0\rho^2} \< -\bar\psi\gamma^0\psi
\left(1-{2\phi^\dagger\phi\0\rho}\right) \>
= -2mr \int {dx\0(1+e^{2mx})^2}
e^{2mx}\left(1-{2e^{2mx}\01+e^{2mx}}
\right)=0,
\ee
whereas the fermion number charge density is given by ${r\0\rho^2}
\bar\psi\gamma^0\psi$ (and in strictly two dimensions
this gives a nonvanishing integral when the fermionic zero mode is inserted).

The final result that we have obtained for the one-loop correction
to the mass of the kink, eq.\ (\ref{M1}), 
and correspondingly for the correction of one
of the central charges, eq.\ (\ref{Z1}), is given by $-m/\pi$.
In the calculation of the previous section where we
considered the central charges we have identified this
contribution
as arising from
a net momentum density associated with fermionic modes
along the domain line (whereas the classical contribution
to the central charge
is a pure surface term). Thus at the quantum level there is a breaking of
parity in the extra regulator dimension which is induced by the kink
background, similar to what occurs in the
minimally susy kink \cite{Rebhan:2002yw}.

Compared to previous calculations of quantum corrections to
two-dimensional susy kinks we have noticed in particular two
new features of the 
$\mathcal N=2$ CP$^1$ model
with twisted mass term: whereas in other $\mathcal N=2$ susy kink models
extended susy leads to a cancellation of the anomalous contributions
\cite{Nastase:1998sy,Rebhan:2002uk},
in the $\mathcal N=2$ CP$^1$ model they add up. Related to this
is the fact that in the $\mathcal N=2$ CP$^1$ model the complex
fermion zero mode has definite chirality with respect to
the domain line employed in our dimensional regularization scheme.
Another noteworthy difference to other susy kinks is the
appearance of fermionic surface terms in the one-loop corrections
to the central charge, cf.\  eq.~(\ref{fermsurfterm}), which
neither occurred in other susy kink models considered so far nor in the
case of 4-dimensional (Coulomb phase) BPS monopoles, 
which with $\mathcal N=2$ also receive anomalous
contributions to their central charge \cite{Rebhan:2004vn}.

To conclude, we have presented an explicit calculation of the
one-loop corrections to both mass and central charge of the
susy kink of the $\mathcal N=2$ nonlinear sigma model with twisted mass
and found agreement with the exact results obtained by
Dorey in Ref.~\cite{Dorey:1998yh}. The nontrivial corrections
have been identified as being associated with an anomalous
contribution to the central charge \cite{Shifman:2006bs}
that in our scheme appears
as parity violation in the higher dimension used to imbed
the susy kink as a domain line, which carries chiral domain
wall fermions.
This mechanism is completely parallel to the anomalous contributions
obtained in the minimally susy kink in 2 dimensions as well
as the $\mathcal N=2$ susy 't Hooft-Polyakov monopole \cite{Rebhan:2004vn}, 
where the
anomalous contribution to the central charge is required for consistency with
the Seiberg-Witten solution. Indeed, as explained in
Ref.~\cite{Shifman:2004dr}, holomorphicity relates the latter to the anomalous
central charge of the nonabelian confined monopoles
appearing in the Higgs phase of $\mathcal N=2$ SU($2$)$\times$U(1) theory, 
whose
effective low energy theory is given by the kinks of
the two-dimensional $\mathcal N=2$
CP$^1$ model with twisted mass.

\acknowledgments

We would like to thank Nick Dorey and Arkady Vainshtein 
for useful discussions.
R.W.\ and P.v.N.\ have been supported by the Austrian
Science Foundation FWF, project no. J2660 and
NSF grant no. PHY-0354776, respectively.

\small

\providecommand{\href}[2]{#2}\begingroup\raggedright\endgroup

\end{document}